\documentclass[conference]{IEEEtran}
\IEEEoverridecommandlockouts
\usepackage{times}
\usepackage{amsmath}
\usepackage{amssymb}
\usepackage{multicol,multirow}
\usepackage[tableposition=top]{caption}
\usepackage{cite}
\usepackage{booktabs}
\usepackage{bigstrut}
\usepackage{floatrow}
\usepackage{rotating}
\usepackage{adjustbox}
\usepackage{amsmath}

\def\BibTeX{{\rm B\kern-.05em{\sc i\kern-.025em b}\kern-.08em
    T\kern-.1667em\lower.7ex\hbox{E}\kern-.125emX}}
\begin{document}
\title{TESL-Net: A Transformer-Enhanced CNN for Accurate Skin Lesion Segmentation}


\author{Shahzaib Iqbal ,~\IEEEmembership{Member,~IEEE,}
  Muhammad Zeeshan,~\IEEEmembership{Member,~IEEE,}
  Mehwish Mehmood,~\IEEEmembership{Member,~IEEE,}\\
        Tariq M. Khan,~\IEEEmembership{Member,~IEEE,}
        and~Imran Razzak,~\IEEEmembership{Member,~IEEE}
\thanks{Shahzaib Iqbal and Muhammad Zeeshan are with the Department of Electrical Engineering, Abasyn University Islamabad Campus(AUIC), Islamabad, Pakistan (email: shahzaib.iqbal91@gmail.com)}

\thanks{Mehwish Mehmood is with the School of Electronics, Electrical Engineering and Computer Science, Queen's University Belfast, United Kingdom (email: mmehmood01@qub.ac.uk)}

\thanks{Tariq M. Khan and Imran Razzak are with the School of Computer Science \& Engineering, UNSW, Sydney, Australia (e-mail: \{tariq.khan, imran.razzak\}@unsw.edu.au)}  
}

\maketitle

\begin{abstract}
Early detection of skin cancer relies on precise segmentation of dermoscopic images of skin lesions. However, this task is challenging due to the irregular shape of the lesion, the lack of sharp borders, and the presence of artefacts such as marker colours and hair follicles. Recent methods for melanoma segmentation are U-Nets and fully connected networks (FCNs). As the depth of these neural network models increases, they can face issues like the vanishing gradient problem and parameter redundancy, potentially leading to a decrease in the Jaccard index of the segmentation model. In this study, we introduced a novel network named TESL-Net for the segmentation of skin lesions. The proposed TESL-Net involves a hybrid network that combines the local features of a CNN encoder-decoder architecture with long-range and temporal dependencies using bi-convolutional long-short-term memory (Bi-ConvLSTM) networks and a Swin transformer. This enables the model to account for the uncertainty of segmentation over time and capture contextual channel relationships in the data. We evaluated the efficacy of TESL-Net in three commonly used datasets (ISIC 2016, ISIC 2017, and ISIC 2018) for the segmentation of skin lesions. The proposed TESL-Net achieves state-of-the-art performance, as evidenced by a significantly elevated Jaccard index demonstrated by empirical results.


\end{abstract}

\vspace{0.5\baselineskip}


\section{Introduction}

\IEEEPARstart{M}{elanoma} is the leading cause of skin cancer-related mortality, presenting a substantial global health concern \cite{siegel2023cancer}. The survival rate for melanoma patients drops below 15\% if the disease is not detected early \cite{wang2019bi}. Therefore, early detection is crucial to reducing mortality rates, with research indicating a 90\% survival rate for patients diagnosed in the early stages. However, differentiating a melanoma lesion from the surrounding healthy skin is challenging. The appearance of the skin can be affected by various factors, including lesion size, hair, reflections, colors, marker colors, textures, shapes, and non-uniform illumination \cite{wu2022fat}.

Dermatoscopy is a non-invasive imaging technique widely used to identify skin lesions and their surrounding areas for the detection and diagnosis of skin cancer \cite{hu2022net}. Manual evaluation of dermoscopic images requires specialised knowledge in dermoscopy and is time-consuming. Even for highly experienced dermatologists, diagnosing skin cancer using only their unaided eye can be imprecise, unreliable, and time-consuming \cite{bi2022hyper}. Traditional image preprocessing techniques struggle with complex tasks due to their reliance on highly customised and precise features and methods \cite{yueksel2009accurate}. To improve the efficacy of lesion analysis and identification, dermatologists have implemented computer-aided diagnostic (CAD) technologies \cite{iqbal2023ldmres, khan2024esdmr}. Precise segmentation is a critical component of any CAD-based diagnostic platform for skin cancer. This process improves the precision and effectiveness of skin lesion segmentation by providing essential quantitative information, including location, size, shape, and other characteristics \cite{cao2022icl}.

\begin{figure}
    \centering
    \resizebox{1.0\textwidth}{!}{%
        \begin{tabular}{ccccc}

            \includegraphics[width=\textwidth]{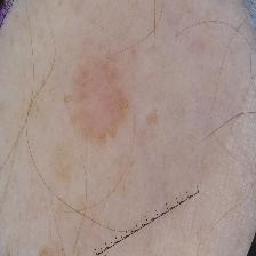} &              \includegraphics[width=\textwidth]{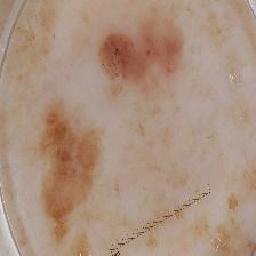} &             \includegraphics[width=\textwidth]{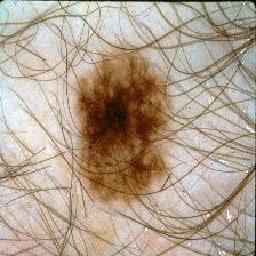} &
            \includegraphics[width=\textwidth]{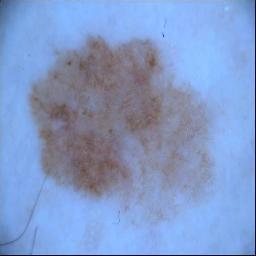} &             \includegraphics[width=\textwidth]{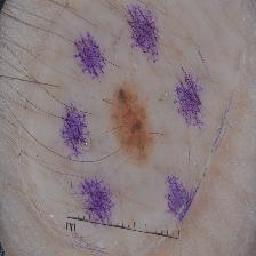} \\
            \\
            
            \includegraphics[width=\textwidth]{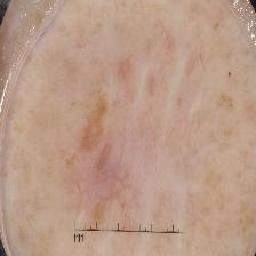} &              \includegraphics[width=\textwidth]{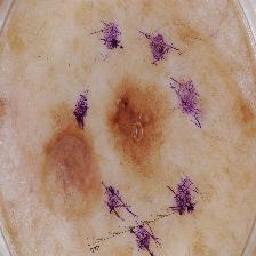} &             \includegraphics[width=\textwidth]{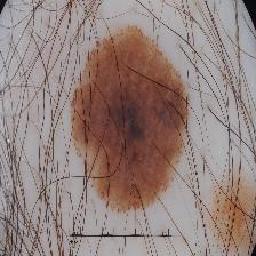} &
            \includegraphics[width=\textwidth]{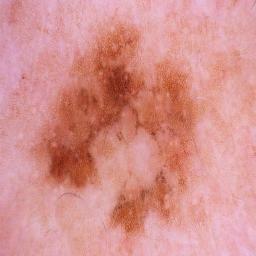} &             \includegraphics[width=\textwidth]{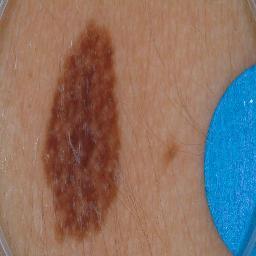}

        \end{tabular}
    }
        \caption{Challenges in skin lesion segmentation.}
    \label{fig:challenges}
\end{figure}

Skin lesion segmentation presents several challenges \cite{minhas2020accurate}. Precisely delineating skin lesions is often difficult due to their irregular and blurry boundaries \cite{khan2022leveraging,khan2022t}. Differentiating between the healthy surrounding skin and a lesion is also frequently challenging. In addition, the varying shapes, sizes, and colours of skin lesions further complicate their characterisation. Interference elements, including blood vessels, ruler traces, hairs, and ink speckles, add to the complexity of segmentation \cite{khan2022mkis, naqvi2023glan}. These challenges are illustrated in Figure \ref{fig:challenges}, where lesions with diverse shapes, sizes and colours, as well as irregular and hazy boundaries, introduce redundancy that reduces performance \cite{iqbal2023ldmres,javed2024advancing}. Low-contrast skin lesions from surrounding healthy tissues and interference elements such as blood vessels, filaments, and ink speckles add noise to images. These factors impede the development of advanced segmentation techniques.
However, skin lesion segmentation presents several challenges \cite{naveed2024pca}. Precisely delineating skin lesions is often difficult due to their irregular and blurry boundaries. Differentiating between the healthy skin surrounding is also often challenging. In addition, the varying shapes, sizes, and colours of skin lesions further complicate their characterisation. Interference elements, including blood vessels, ruler traces, hairs, and ink speckles, add to the complexity of segmentation. These challenges are illustrated in Figure \ref{fig:challenges}, where lesions with diverse shapes, sizes and colours, as well as irregular and hazy boundaries, introduce redundancy that reduces performance. Low-contrast skin lesions from surrounding healthy tissue, and interference features such as blood vessels, filaments, and ink speckles add noise to images. These factors impede the development of advanced segmentation techniques.

\begin{figure*}[h]
    \centering
    \includegraphics[width=0.75\textwidth]{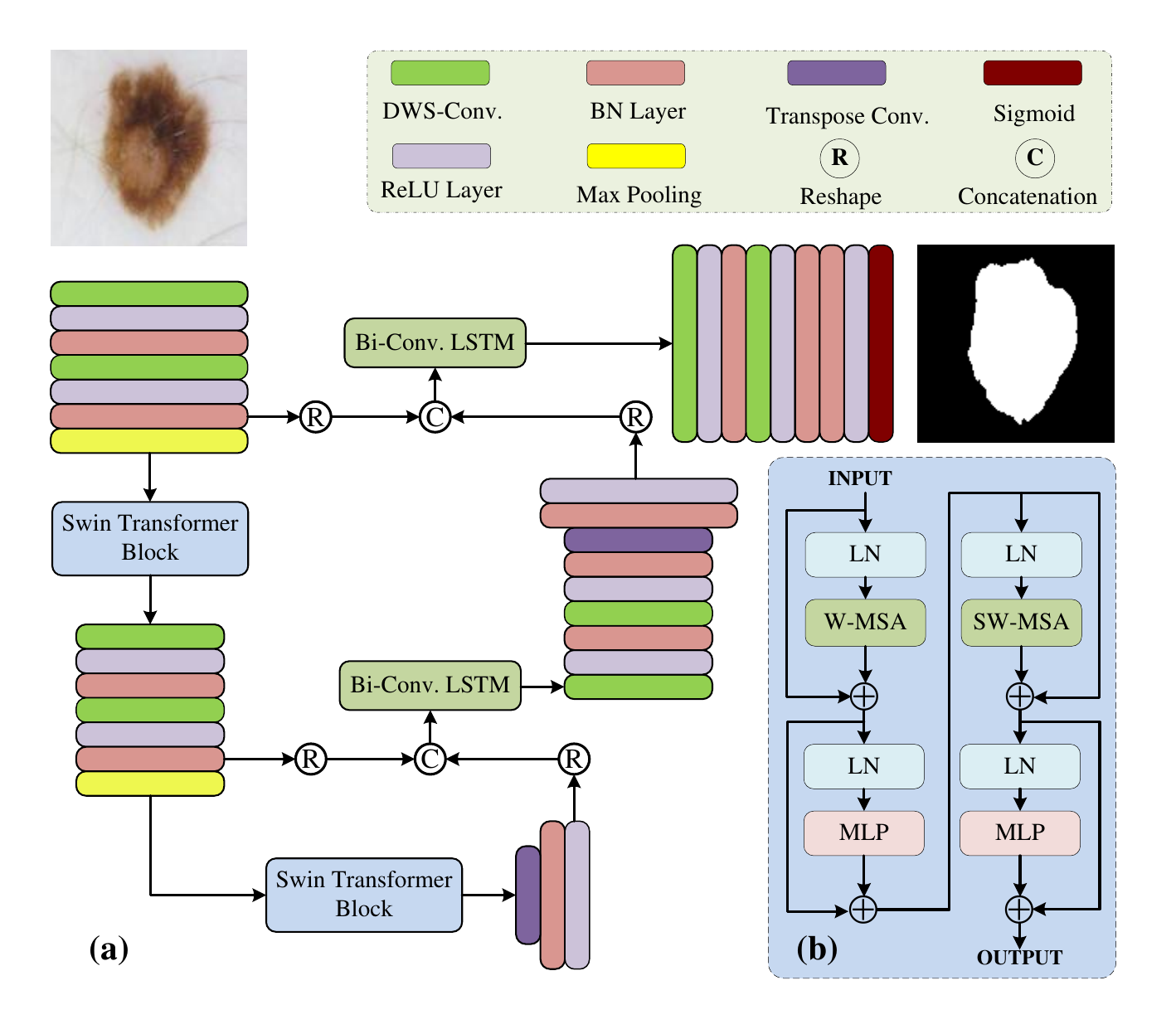}
    \caption{Schematic of the proposed method. (a) Block diagram of the proposed TESL-Net, (b) Swin Transformer Block.}
    \label{fig:model}
\end{figure*}

For the task of segmenting skin lesions, a variety of convolutional neural network (CNN) techniques, as well as attention-based approaches, have been explored. Bi \textit{et al.} designed a network that extracts contextual and hierarchical information by integrating the combination of pyramidal features, residual connections, and dilated convolution \cite{bi2019step}. Tang \textit{et al.} introduced DeepLabV3+, a CNN architecture that incorporates an advanced spatial pyramid pooling module to extract multi-scale features \cite{tang2019efficient}. Another notable example is the U-Net architecture \cite{ronneberger2015u}, which has become the industry standard for medical image segmentation, including skin lesions. The advent of deep learning has significantly improved the analysis of biological data and image segmentation \cite{mahmud2021deep}. By effectively utilizing relevant features, deep learning methods outperform traditional methods in skin lesion segmentation. Segmentation performance has been further enhanced by improvements to the encoder-decoder architecture, including the implementation of efficient feature map learning procedures \cite{dai2022ms}.

Segmentation model training can be enhanced by data augmentation techniques, such as rotation, scaling, and flipping, which increase the scale and diversity of datasets \cite{iqbal2022recent}. To achieve optimal results, it is essential to carefully regulate the selection and extent of augmentation. Deep neural network models with numerous layers may encounter issues such as parameter redundancy and vanishing gradients. To address these challenges and achieve precise skin lesion segmentation, we have developed a  transformer-enhanced CNN, TESL-Net. Our proposed technique ensures accurate segmentation of skin lesions while maintaining a model architecture.

\section{Related Work}
\label{sec:Related Work}



Numerous segmentation techniques are emphasised in the literature for the segmentation of skin lesions, including morphological operations \cite{817154}, thresholding approaches \cite{emre2013lesion}, gradient vector flow \cite{erkol2005automatic}, and growth of the region \cite{7004778}. These conventional methods typically involve threshold setting, feature selection, and image pre-processing. The emergence of deep learning has significantly advanced segmentation techniques, particularly with CNNS. Yuan and Lo developed an improved CNN for skin lesion segmentation \cite{yuan2017improving}. Furthermore, studies have also used multiscale connection blocks instead of traditional skip connections to capture features at both the low- and the high-level more effectively \cite{8848426}.
Hasan \textit{et al.} proposed a dermoscopic skin network that employs depthwise separable convolution to reduce the number of trainable parameters \cite{HASAN2020103738}. Abhishek \textit{et al.} devised a novel deep semantic segmentation method that takes advantage of data from multiple colour bands \cite{abhishek2020illumination}. To analyse the boundaries and contextual relationships of target objects, the DAGAN authors implemented a dual discriminator network \cite{LEI2020101716}.

Attention mechanisms have been extensively implemented in CNNs to facilitate various tasks, including semantic segmentation, identification, classification, and machine translation. This approach enables models to focus on the most relevant features, thus reducing computational demands by weighting the features to emphasise pertinent information and suppress irrelevant data \cite{naveed2024pca,wang2020attentive}. To optimise skin lesion segmentation, Chen \textit{et al.} integrated self-attention within codec components \cite{chen2021transattunet}. Zhang \textit{et al.} developed an attention-directed filter within a U-shaped framework to convey spatial features for image segmentation \cite{zhang2019attention}. A channel attention strategy was implemented to enhance skin lesion segmentation in a generative adversarial network (GAN) \cite{8832175}.  CS2-Net enhanced feature extraction by implementing dual attention strategies in both the spatial and channel domains \cite{mou2021cs2}. The AS-Net further improved segmentation performance by integrating spatial and channel attention techniques \cite{HU2022117112}. Furthermore, networks like MFSNet \cite{BASAK2022108673} increased segmentation efficiency by incorporating multiscale concepts with attention mechanisms.

\begin{table*}[h]
  \centering
  \caption{Details of the skin lesion image datasets used for TESL-Net evaluation.}
  \resizebox{1\textwidth}{!}{%
    \begin{tabular}{lccccccc}
    \toprule
    \multirow{2}[4]{*}{\textbf{Dataset}} & \multicolumn{4}{c}{\textbf{Image Count}} & \multirow{2}[4]{*}{\textbf{Image Resolution Range}} & \multirow{2}[4]{*}{\textbf{Format}} & \multirow{2}[4]{*}{\textbf{Resized to}} \\
\cmidrule{2-5}          & \textbf{Training} & \textbf{Validation} & \textbf{Testing} & \textbf{Total} &       &       &  \\
    \midrule
    ISIC 2016 \cite{gutman2016skin} & 900   & -     & 379   & 1279  & 679x453 - 6748x4499 & .jpeg & \multirow{3}[2]{*}{256x256} \\
    ISIC 2017 \cite{codella2018skin}  & 2000  & 150   & 600   & 2750  & 679×453 - 6748×4499 & .jpeg &  \\
    ISIC 2018 \cite{codella2019skin}& 2594  & -     & 1000  & 3594  & 679×453 - 6748×4499 & .jpeg &  \\
    \bottomrule
    \end{tabular}%
    }
  \label{tab:database}%
\end{table*}%

\section{Proposed Method}\label{method}

The proposed network uses Bidirectional Convolutional Long-Short-Term Memory (Bi-ConvLSTM) layers and spin transformer blocks to segment skin lesions.

\subsection{TESL-Net}
The architecture of the proposed TESL-Net takes RGB images along with their corresponding masks as input. At the encoder stage, two consecutive blocks of depth-wise convolution followed by the activation function and batch normalisation layer are applied. After that, a max pooling layer is employed to reduce the spatial dimensions of the features. Once the size of the feature maps is reduced a Swin transformer block is used to extract and refine the feature information patch-wise. The same operations are again applied on the feature maps by increasing the depth twice. It is important to mention that the proposed TESL-Net uses two max-pooling layers at the encoder stage so that the spatial information especially at the boundary can be kept intact. At the decoder stage of the proposed TESL-Net transposed convolution is used to upsample the feature maps followed by ReLU and batch normalization operations. Once the spatial dimensions are increased, the Bi-ConvLSTM is used between the encoder-decoder blocks to capture short-term details and long-term dependencies of the feature information. Two consecutive blocks of depth-wise convolution followed by the activation function and batch normalization layer are then employed to reconstruct the extracted feature information. The same operations are again applied on the feature maps by reducing the channel depth twice. Finally, the sigmoid layer is employed to predict the binary masks. The mathematical description of the proposed TESL-Net is as follows:
Let $I$ be the RGB image of size $(H\times W \times 3)$ given to the input layer. The Depthwise Separable Convolution (DWS-Conv) is applied to the input image followed by batch normalization $\beta_{n}$ and the ReLU (Rectified Linear Unit) activation function $\Re$ that helps in dealing with overfitting and introduces non-linearity into the network. The output is defined as:
\begin{equation}
 \texttt{C}_{1}=\beta_{n} \left [ \Re (\texttt{DWS-Conv}(I)_{i,j,c})\right ] )
 \end{equation}
 The resulting feature map is again fed into $DWS-Conv$ followed by ReLU and BN. 
 \begin{equation}
 \texttt{C}_{2}=\beta_{n} \left [ \Re (\texttt{DWS-Conv}(C_{1})\right ] )
 \end{equation}
The feature map \( C_2 \) is passed through the max-pooling operation and then processed by the Swin Transformer Block, which captures long-range dependencies and context information. 
\begin{equation}
 \texttt{ST}=\texttt{Swin-block}(m_{p} \left (C_{2})\right ] ))
 \end{equation}
A similar process is applied to the resulting feature map in the second convolutional block. It is defined as:
\begin{equation}
 \texttt{C}_{3}=\beta_{n} \left [ \Re (\texttt{DWS}(\texttt{ST})\right ] )
 \end{equation}
 \begin{equation}
 \texttt{C}_{4}=\beta_{n} \left [ \Re (\texttt{DWS}(\texttt{C}_{3})\right ] )
 \end{equation}

Subsequently, a transposed convolution operation is applied to the encoder generated feature map to up-sample the feature maps, followed by a ReLU activation layer $\Re$ and $\beta_{n}$.
The feature map \( C_2 \) and \( C_4 \) from the encoder are reshaped and concatenated with the corresponding reshaped feature maps from the decoder. The outputs are then fed into the corresponding Bidirectional Convolutional LSTM (Bi-ConvLSTM) to capture temporal dependencies in both forward and backward directions. 
The forward ConvLSTM is defined as:

\begin{equation}
\begin{aligned}
i_t &= \sigma(W_i \cdot [h_{t-1}, x_t] + b_i) \\
o_t &= \sigma(W_o \cdot [h_{t-1}, x_t] + b_o) \\
\tilde{C_t} &= \tanh(W_C \cdot [h_{t-1}, x_t] + b_C) \\
f_t &= \sigma(W_f \cdot [h_{t-1}, x_t] + b_f) \\
C_t &= f_t \odot C_{t-1} + i_t \odot \tilde{C_t} \\
h_t &= o_t \odot \tanh(C_t)
\end{aligned}
\end{equation}
where \( f_t \), \( i_t \), and \( o_t \) are the forget, input, and output gates, \( \tilde{C_t} \) is the cell input activation, \( C_t \) is the cell state, and \( h_t \) is the hidden state.

The backward ConvLSTM operates similarly but in reverse temporal order. The Bi-ConvLSTM combines the forward and backward ConvLSTM outputs:
\begin{equation}
\texttt{Bi-ConvLSTM}(x) = [h_t^{\texttt{forward}}, h_t^{\texttt{backward}}]
\end{equation}

The respective outcomes of \texttt{Bi-ConvLSTM} are processed through \texttt{DWS-Conv}, $\Re$ and $\beta_{n}$ along with transposed convolution.
The final predicted mask is computed by applying a sigmoid to the last layer for binary segmentation.

\begin{figure*}[h]
    \centering
     \resizebox{1.0\textwidth}{!}{%
     
     \begin{tabular}{cccccccccccccccccccc}
        \includegraphics[width=\textwidth]{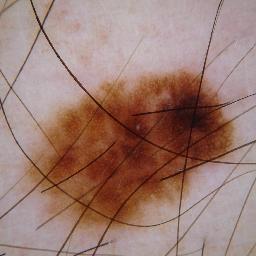} && \includegraphics[width=\textwidth]{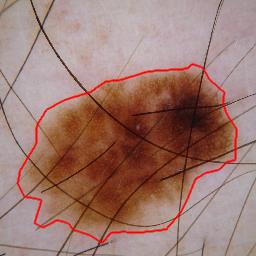} && \includegraphics[width=\textwidth]{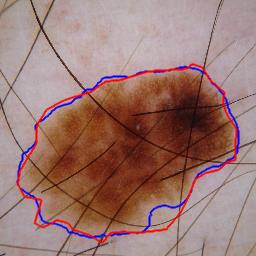} && \includegraphics[width=\textwidth]{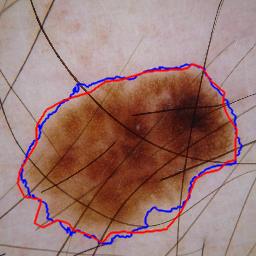} && \includegraphics[width=\textwidth]{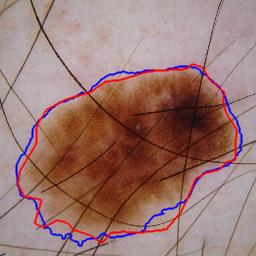} && \includegraphics[width=\textwidth]{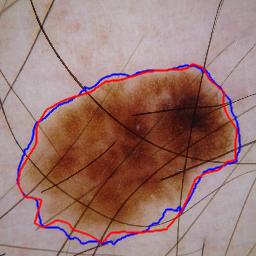} && \includegraphics[width=\textwidth]{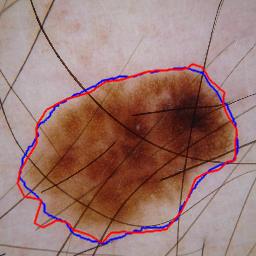} && \includegraphics[width=\textwidth]{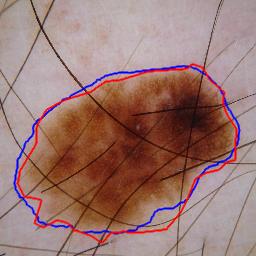} && \includegraphics[width=\textwidth]{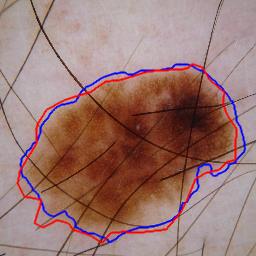} && \includegraphics[width=\textwidth]{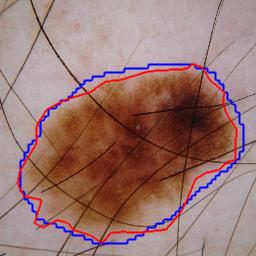} \\

\\
        \includegraphics[width=\textwidth]{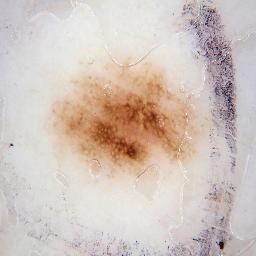} && \includegraphics[width=\textwidth]{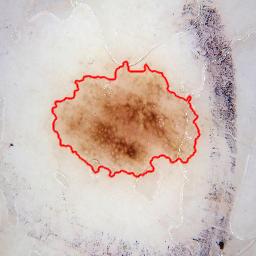} && \includegraphics[width=\textwidth]{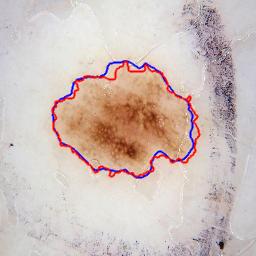} && \includegraphics[width=\textwidth]{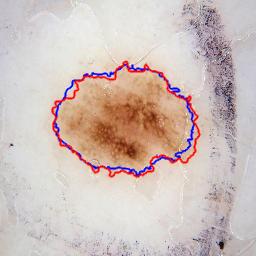} && \includegraphics[width=\textwidth]{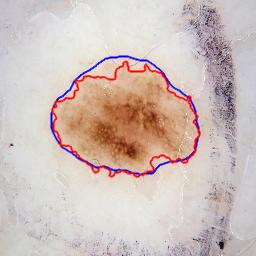} && \includegraphics[width=\textwidth]{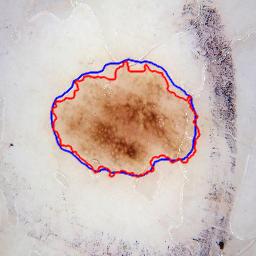} && \includegraphics[width=\textwidth]{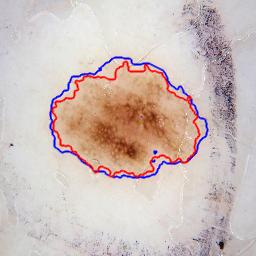} && \includegraphics[width=\textwidth]{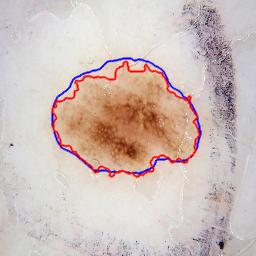} && \includegraphics[width=\textwidth]{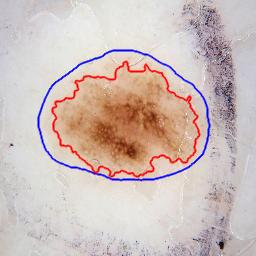} && \includegraphics[width=\textwidth]{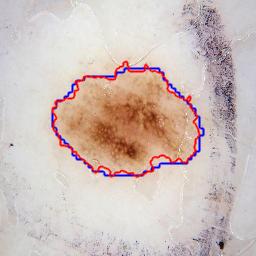} \\
                \\
        \end{tabular}
    }
                 \centering
            \resizebox{1.0\textwidth}{!}{%
            \begin{tabular}{cccccccccccccccccccc}
              \hspace{5mm}  Image  &&  \hspace{5mm}   GT && \hspace{5mm}     Proposed &&  Swin-Unet \cite{cao2023swin} && U-Net \cite{ronneberger2015u} && ARU-GD \cite{Maji2022} && DAGAN \cite{LEI2020101716} && UNet++ \cite{Zhou2018} && RA-Net \cite{naveed2024ra} && FAT-Net \cite{WU2022102327} \\
                    
        \end{tabular}
    }
    \caption{Visual performance comparison of the proposed TESL-Net on ISIC 2016 \cite{gutman2016skin} dataset.}
    \label{fig:Vis_ISIC2016}
\end{figure*}

\subsection{Swin Transformer Block}


In contrast to the traditional multi-head self-attention (MSA) module, the Swin Transformer block is designed using shifted windows. (Fig. (\ref{fig:model})-b) illustrates two consecutive Swin Transformer blocks. Each Swin Transformer block includes a residual connection, a multi-head self-attention module, a LayerNorm (LN) layer, and a two-layer MLP with GELU activation. The two successive transformer blocks utilise the window-based multihead self-attention module (W-MSA) and the shifted window-based multihead self-attention module (SW-MSA), respectively. The following equations describe the formulation of continuous Swin Transformer blocks through this window partitioning mechanism:

\begin{equation}
{\hat{x}^{l}}=\texttt{W-MSA}\left ( \texttt{LN}\left ( z^{l-1} \right ) \right ) +z^{l-1},
    \label{eq:1}
\end{equation}

\begin{equation}
{z^{l}}=\texttt{MLP}\left ( \texttt{LN}\left ( \hat{x}^l \right ) \right ) + \hat{x}^l,
    \label{eq:2}
\end{equation}

\begin{equation}
\hat{x}^{l+1} =\texttt{SW} - \texttt{MSA}(\texttt{LN}(x^l)) + z^l
    \label{eq:3}
\end{equation}

\begin{equation}
x^{l+1} = \texttt{MLP}(\texttt{LN}(\hat{x}^{l+1}))+\hat{x}^{l+1}
    \label{eq:4}
\end{equation}

where $\hat{x}^l and z^l $ represents the output of the module \texttt{ (S) W-MSA} and the \texttt{MLP} module of $l^th$ block.
Self Attention is computed as follows:
\begin{equation}
\texttt{Attention} (Q,K,V) = \texttt{SoftMax}\left ( \frac{QK^T}{\sqrt{d}}+B \right )V,
    \label{eq:5}
\end{equation}
where $(Q,K,V \epsilon \mathbb{R}^{{M}^2} \times d)$ denotes query, key and value Matrices. $M^2$ and $d$ represent several patches in the window and the dimension of the key or query, respectively. where $B$ is the value taken from the bias matrix $\hat{B \epsilon \mathbb{R}^{(2M-1) \times(2M+1)}}$

\section{Results and Discussion}\label{experimentalResults}

TESL-Net was evaluated against several SOTA segmentation networks. This section provides an overview of the datasets used, the evaluation criteria, the experimental setup, and the comparative experiments.

\subsection{Datasets}

The proposed TESL-Net model was evaluated in three challenging benchmark datasets (Table \ref{tab:database}), namely ISIC 2016 \cite{gutman2016skin}, ISIC 2017 \cite{codella2018skin} and ISIC 2018 \cite{codella2019skin} for the segmentation of skin lesions in optical images. All datasets are publicly available and provide GT masks for the evaluation of image segmentation methods.



\begin{table}[!b]
  \centering
  \caption{Quantitative performance comparison of TESL-Net with various SOTA methods on the ISIC2016 skin lesion dataset.}
    \resizebox{1\textwidth}{!}{%
    \begin{tabular}{lccccc}
    \toprule
    \multirow{2}[4]{*}{\textbf{Method}} & \multicolumn{5}{c}{\textbf{Performance Measures  (\%)}} \\
    \cmidrule{2-6} & \bf IoU & \bf Dice & \bf Acc & \bf Se & \bf Sp \\
    \midrule
    ARU-GD \cite{Att_resUnet} & 85.12 & 90.83 & 94.38 & 89.86 & 94.65 \\
    BCDU-Net \cite{azad2019bi} & 83.43 & 80.95 & 91.78 & 78.11 & 96.20 \\
    CPFNet \cite{9049412} & 83.81 & 90.23 & 95.09 & 92.11 & 95.91 \\
    DAGAN \cite{LEI2020101716}  & 84.42 & 90.85 & 95.82 & 92.28 & 95.68 \\
    FAT-Net \cite{WU2022102327} & 85.30 & 91.59 & 96.04 & 92.59 & 96.02 \\
    RA-Net \cite{naveed2024ra} & 87.40 & 92.94 & 96.70 & 92.27 & 96.79 \\
    
    Separable-Unet \cite{TANG2019289} & 84.27 & 89.95 & 95.67 & 93.14 & 94.68 \\
    Swin-Unet \cite{cao2023swin} & 87.60 & 88.94 & 96.00 & 92.27 & 95.79 \\     
    U-Net \cite{ronneberger2015u}  & 81.38 & 88.24 & 93.31 & 87.28 & 92.88 \\
    UNet++ \cite{zhou2019unet++} & 82.81 & 89.19 & 93.88 & 88.78 & 93.52 \\

    \midrule
    \textbf{Proposed Method} & \textbf{89.51} & \textbf{93.43} & \textbf{96.40} & \textbf{94.55} & \textbf{97.02} \\
    \bottomrule
    \end{tabular}%
    }
  \label{tab:ISIC2016}%
\end{table}%

\begin{figure*}[h]
    \centering
    \resizebox{1.0\textwidth}{!}{%
        \begin{tabular}{cccccccccccccccccccc}

            \includegraphics[width=\textwidth]{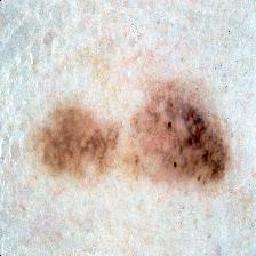} && \includegraphics[width=\textwidth]{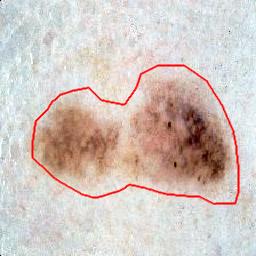} && \includegraphics[width=\textwidth]{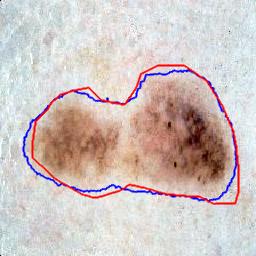} && \includegraphics[width=\textwidth]{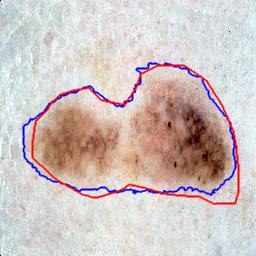} && \includegraphics[width=\textwidth]{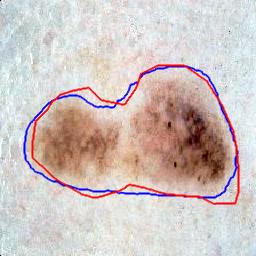} && \includegraphics[width=\textwidth]{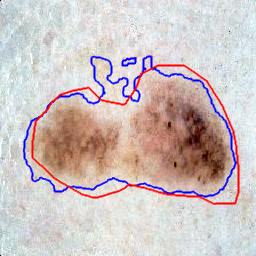} && \includegraphics[width=\textwidth]{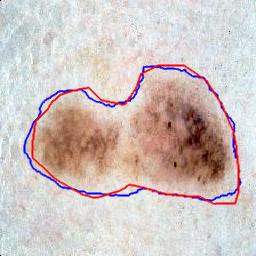} && \includegraphics[width=\textwidth]{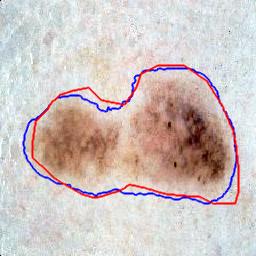} && \includegraphics[width=\textwidth]{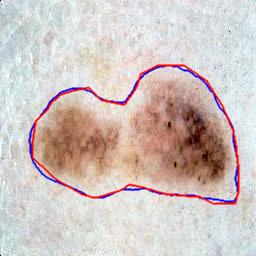} && \includegraphics[width=\textwidth]{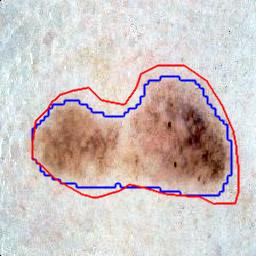} \\
            \\
            \includegraphics[width=\textwidth]{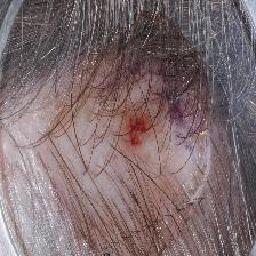} && \includegraphics[width=\textwidth]{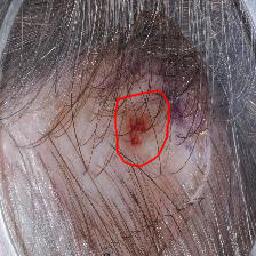} && \includegraphics[width=\textwidth]{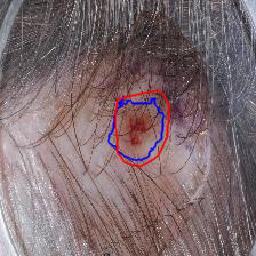} && \includegraphics[width=\textwidth]{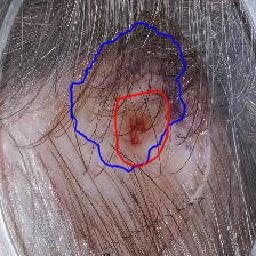} && \includegraphics[width=\textwidth]{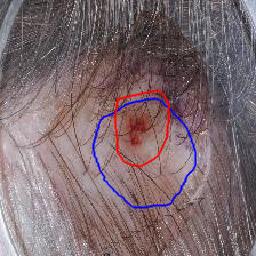} && \includegraphics[width=\textwidth]{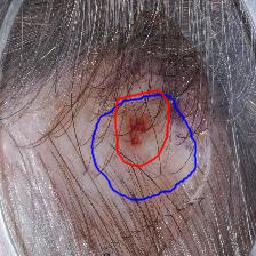} && \includegraphics[width=\textwidth]{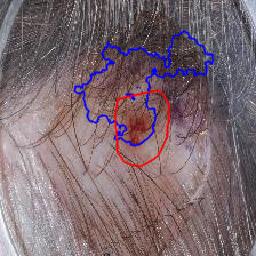} && \includegraphics[width=\textwidth]{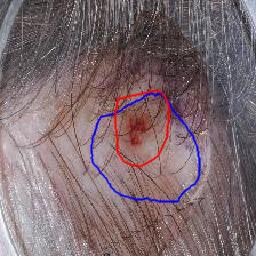} && \includegraphics[width=\textwidth]{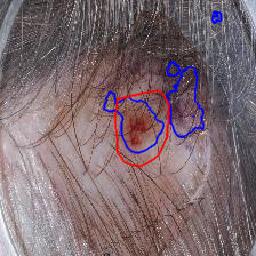} && \includegraphics[width=\textwidth]{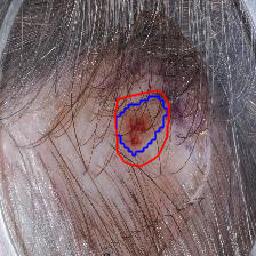} \\
                 \end{tabular}
    }         
           \centering
            \resizebox{1.0\textwidth}{!}{%
            \begin{tabular}{cccccccccccccccccccc}
              \hspace{5mm}  Image  &&  \hspace{5mm}   GT && \hspace{5mm}     Proposed &&  Swin-Unet \cite{cao2023swin} && U-Net \cite{ronneberger2015u} && ARU-GD \cite{Maji2022} && DAGAN \cite{LEI2020101716} && UNet++ \cite{Zhou2018} && RA-Net \cite{naveed2024ra} && FAT-Net \cite{WU2022102327} \\
                    
        \end{tabular}
    }
        \caption{Visual performance comparison of the proposed TESL-Net on ISIC 2017 \cite{codella2018skin} dataset.}
    \label{fig:Vis_ISIC2017}
\end{figure*}

\begin{figure*}[h]
    \centering
     \resizebox{1.0\textwidth}{!}{%
     
     \begin{tabular}{cccccccccccccccccccc}
       
        \includegraphics[width=\textwidth]{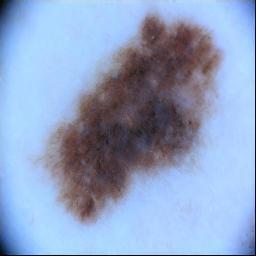} && \includegraphics[width=\textwidth]{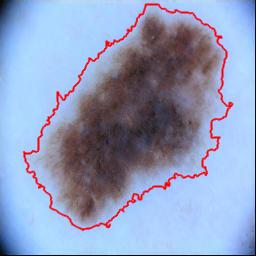} && \includegraphics[width=\textwidth]{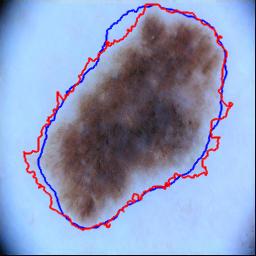} && \includegraphics[width=\textwidth]{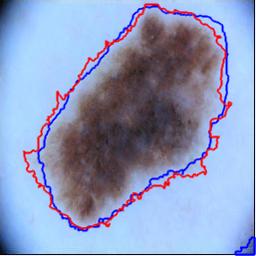} && \includegraphics[width=\textwidth]{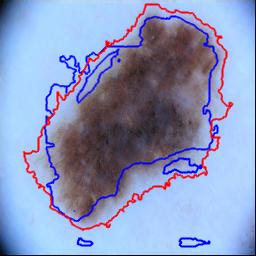} && \includegraphics[width=\textwidth]{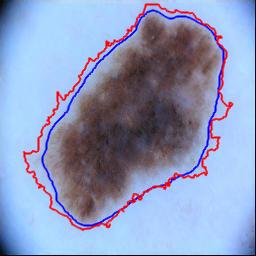} && \includegraphics[width=\textwidth]{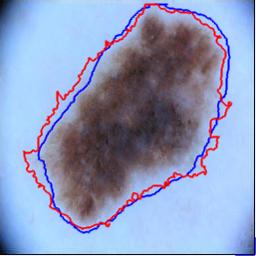} && \includegraphics[width=\textwidth]{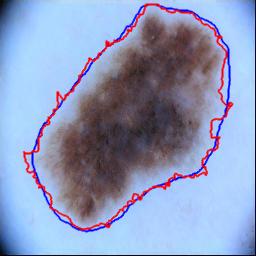} && \includegraphics[width=\textwidth]{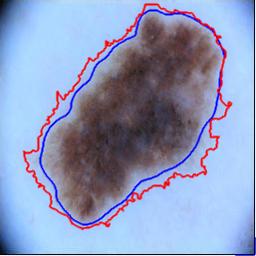} && \includegraphics[width=\textwidth]{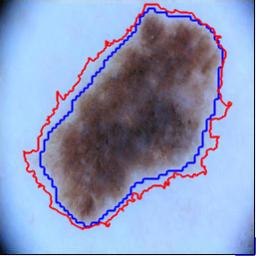} \\
\\

       \includegraphics[width=\textwidth]{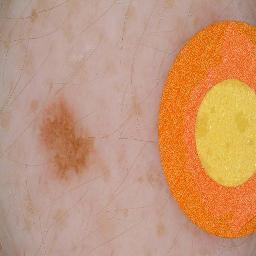} && \includegraphics[width=\textwidth]{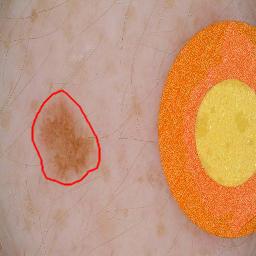} && \includegraphics[width=\textwidth]{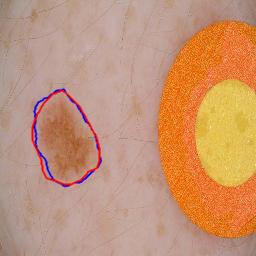} && \includegraphics[width=\textwidth]{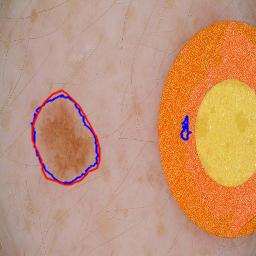} && \includegraphics[width=\textwidth]{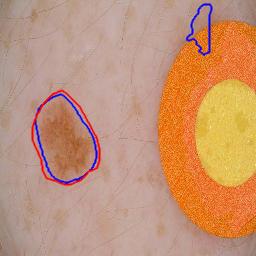} && \includegraphics[width=\textwidth]{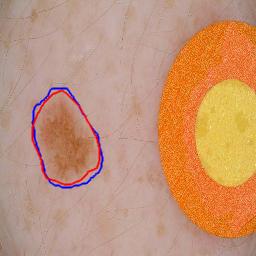} && \includegraphics[width=\textwidth]{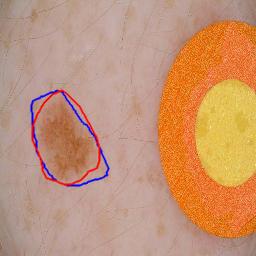} && \includegraphics[width=\textwidth]{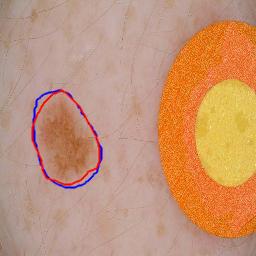} && \includegraphics[width=\textwidth]{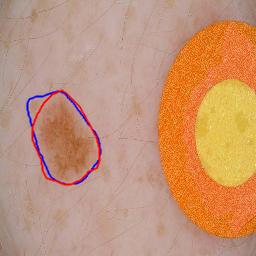} && \includegraphics[width=\textwidth]{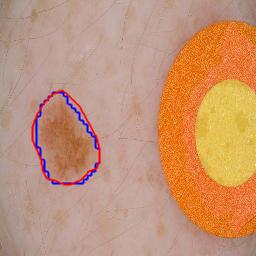} \\
                \\
        \end{tabular}
    }
            \centering
            \resizebox{1.0\textwidth}{!}{%
            \begin{tabular}{cccccccccccccccccccc}
              \hspace{5mm}  Image  &&  \hspace{5mm}   GT && \hspace{5mm}     Proposed &&  Swin-Unet \cite{cao2023swin} && U-Net \cite{ronneberger2015u} && ARU-GD \cite{Maji2022} && DAGAN \cite{LEI2020101716} && UNet++ \cite{Zhou2018} && RA-Net \cite{naveed2024ra} && FAT-Net \cite{WU2022102327} \\
                    
        \end{tabular}
    }
    \caption{Visual performance comparison of the proposed TESL-Net on ISIC 2018 \cite{codella2019skin} dataset.}
    \label{fig:Vis_ISIC2018}
\end{figure*}
\subsection{Evaluation Criteria}

Performance evaluation of the proposed LSSF-Net is performed using five evaluation metrics recommended by the ISIC challenge leaderboard, including accuracy, Jaccard index (IOU), Dice coefficient, sensitivity, and specificity. These metrics are calculated using counts of true negatives (TN), true positives (TP), false negatives (FN), and false positives (FP) derived from predictions as given in equations (\ref{P1}-\ref{P5}).

\begin{equation}
\mathrm{Accuracy (Acc)} = \frac{{T_{P}+T_{N}}}{{T_{P}+T_{N}+F_{P}+F_{N}}}
\label{P1}
\end{equation}
\begin{equation}
\mathrm{Sensitivity (Se)} = \frac{{T_{P}}}{{T_{P} + F_{N}}}
\label{P2}
\end{equation}
\begin{equation}
\mathrm{IoU} = \frac{{T_{P}}}{{T_{P}+ F_{P} +F_{N}}}
\label{P3}
\end{equation}
\begin{equation}
\mathrm{Dice} = \frac{{2\times T_{P}}}{{2\times T_{P}+ F_{P} +F_{N}}}
\label{P4}
\end{equation}
\begin{equation}
\mathrm{Specificity (Sp)} = \frac{{T_{N}}}{{T_{N} + F_{P}}}
\label{P5}
\end{equation}

\subsection{Experimental Setup}
We assess the effectiveness of the proposed methodology using established benchmark datasets. All data sets were standardised to dimensions of $256\times 256$ pixels to ensure uniformity. A subset comprising 20\% of the training data was segregated for validation purposes. Segmentation models were trained under various loss function configurations, using the optimiser (Adam) over 10 epochs. Initially, a learning rate of 0.001 was set, with a scheduled reduction by a factor of 0.25 every four epochs in the absence of observable improvements on the validation set. In addition, an early stop mechanism was employed to counteract overfitting. In particular, our approach achieved superior performance metrics, exceeding existing benchmarks even without employing data augmentation. The framework was implemented using Keras with TensorFlow backend, and all computations were performed on a NVIDIA K80 GPU.

\subsection{Comparisons with SOTA Methods}

We compared our proposed approach with ten cutting-edge methods including ARU-GD \cite{Maji2022}, BCD-UNet \cite{azad2019bi}, CPFNet \cite{9049412}, DAGAN \cite{LEI2020101716}, FAT-Net \cite{WU2022102327}, RA-Net \cite{naveed2024ra}, Separable-Unet \cite{TANG2019289}, Swin-Unet \cite{cao2023swin}, U-Net \cite{ronneberger2015u}, and UNet++ \cite{Zhou2018}. 

Statistical comparison findings with SOTA techniques in the ISIC 2016 dataset are presented in Table ~\ref{tab:ISIC2016}. Our method consistently outperformed the most advanced techniques in the ISIC 2016 dataset in every metric. Specifically, TESL-Net achieved a Jaccard index (IOU) score that ranged from 2. 11\% to 8. 13\% higher compared to SOTA methods. Our technique demonstrates superior performance in all evaluation criteria. Comparisons of visual results showing various challenges in skin lesion segmentation, such as artefacts, hair, irregular morphologies, and multiple lesions, are presented in Figure~\ref{fig:Vis_ISIC2016}. The TESL-Net method achieved SOTA segmentation results in the test data, effectively handling skin lesions with irregular shapes and varying sizes.


Ten cutting-edge techniques were used to illustrate the statistical comparison findings in the ISIC 2017 dataset, as presented in Table \ref{tab:ISIC2017}. TESL-Net achieved a Jaccard index (IOU) score of 2. 02\% to 11. 22\% higher than the SOTA methods. The visual results showing various challenges in skin lesion segmentation, such as irregular morphologies, hair, and artefacts, are shown in Figure \ref{fig:Vis_ISIC2017}. It is evident that our TESL-Net consistently produces SOTA segmentation results in test data, effectively handling skin lesions with unusual shapes and variable sizes.


\begin{table}[!t]
  \centering
  \caption{Quantitative performance comparison of TESL-Net with various SOTA methods on the ISIC2017 skin lesion dataset.}
      \resizebox{1\textwidth}{!}{%
    \begin{tabular}{lccccc}
    \toprule
    \multirow{2}[4]{*}{\textbf{Method}} & \multicolumn{5}{c}{\textbf{Performance Measures  (\%)}} \\
    \cmidrule{2-6} & \bf IoU & \bf Dice & \bf Acc & \bf Se & \bf Sp \\
    \midrule
    ARU-GD \cite{Att_resUnet} & 80.77 & 87.89 & 93.88 & 88.31 & 96.31 \\
    AS-Net \cite{HU2022117112}  & 80.51 & 88.07 & 94.66 & 89.92 & 95.72 \\
    BCDU-Net \cite{azad2019bi} & 79.20 & 78.11 & 91.63 & 76.46 & 97.09 \\
    DAGAN \cite{LEI2020101716}  & 75.94 & 84.25 & 93.26 & 83.63 & 97.25 \\
    FAT-Net \cite{WU2022102327} & 76.53 & 85.00 & 93.26 & 83.92 & 97.25 \\
    RA-Net \cite{naveed2024ra} & 84.89 & \textbf{90.99} & 95.76 & 91.06 & 96.05 \\
    SLT-Net \cite{FENG2022105942} & 79.87 & 67.90 & -     & 73.63 & 97.27 \\
    Swin-Unet \cite{cao2023swin} & 80.89 & 81.99 & 94.76 & 88.06 & 96.05 \\
    U-Net \cite{ronneberger2015u}  & 75.69 & 84.12 & 93.29 & 84.30 & 93.41 \\
    UNet++ \cite{zhou2019unet++} & 78.58 & 86.35 & 93.73 & 87.13 & 94.41 \\
    
    \midrule
    \textbf{Proposed Method} &   \textbf{86.91}    &    90.09   &   \textbf{95.80}    &   \textbf{91.10}    &  \textbf{97.29}\\
    \bottomrule
    \end{tabular}%
    }
  \label{tab:ISIC2017}%
\end{table}%

Eleven cutting-edge techniques are used to present the statistical comparison findings in Table~\ref{tab:ISIC2018} on the ISIC 2018. In terms of the Jaccard index (IOU), TESL-Net achieved a score of 2.22\%--10.47\% higher than the SOTA methods described. In the same vein, we also obtained visual results for various skin lesion challenges, including the presence of artefacts, low contrast, irregular morphologies, and small lesions. The visual results of numerous skin lesion challenges are illustrated in Figure~\ref{fig:Vis_ISIC2018}. Even for skin lesions with unusual shapes and variable sizes, our method generates SOTA segmentation results on test data.


\begin{table}[htbp]
  \centering
  \caption{Quantitative performance comparison of TESL-Net with various SOTA methods on the ISIC2018 skin lesion dataset.}
      \resizebox{1\textwidth}{!}{%
      \begin{tabular}{lccccc}
    \toprule
    \multirow{2}[4]{*}{\textbf{Method}} & \multicolumn{5}{c}{\textbf{Performance Measures (\%)}} \\
    \cmidrule{2-6} & \bf IoU & \bf Dice & \bf Acc & \bf Se & \bf Sp \\
    \midrule
    ARU-GD \cite{Att_resUnet} & 84.55 & 89.16 & 94.23 & 91.42 & 96.81 \\
    AS-Net \cite{HU2022117112}  & 83.09 & 89.55 & 95.68 & 93.06 & 94.69 \\
    BCDU-Net \cite{azad2019bi} & 81.10 & 85.10 & 93.70 & 78.50 & 98.20 \\
    DAGAN \cite{LEI2020101716}  & 81.13 & 88.07 & 93.24 & 90.72 & 95.88 \\
    FAT-Net \cite{WU2022102327} & 82.02 & 89.03 & 95.78 & 91.00 & 96.99 \\
    ICL-Net \cite{cao2022icl}   & 83.76 & 90.41 & \textbf{97.24} & 91.66 & 98.63 \\
    RA-Net \cite{naveed2024ra} & 88.34 & 93.25 & 95.84 & 93.63 & 94.16 \\
    Swin-Unet \cite{cao2023swin} & 82.79 & 88.98 & 96.83 & 90.10 & 97.16 \\
    SLT-Net \cite{FENG2022105942} & 71.51 & 82.85 & -     & 78.85 & \textbf{99.35} \\
    U-Net \cite{ronneberger2015u}  & 80.09 & 86.64 & 92.52 & 85.22 & 92.09 \\
    UNet++ \cite{zhou2019unet++} & 81.62 & 87.32 & 93.72 & 88.70 & 93.96 \\

    \midrule
    \textbf{Proposed Method} &   \textbf{90.56}    &  \textbf{94.22}     &    96.23   &     \textbf{95.02}  & 97.21 \\
    \bottomrule
    \end{tabular}%
    }
  \label{tab:ISIC2018}%
\end{table}%

\section{Conclusions}
We have developed TESL-Net, a novel and effective methodology for accurate skin lesion segmentation aimed at overcoming challenges in this field. Unlike traditional CNN-based encoder-decoders, TESL-Net utilises Swin transformer blocks in the encoder to efficiently extract contextual information from skin-lesion images globally. Further refinement of feature extraction is achieved by integrating a Bi-ConvLSTM module into skip connections. When evaluated on three publicly available benchmark datasets for skin lesion segmentation, TESL-Net outperformed a variety of SOTA methods. Despite its exceptional performance, we have identified areas for further improvement. We propose employing semi-supervised strategies to reduce data requirements for training by incorporating paired and unpaired data. TESL-Net is suitable not only for immediate medical imaging applications in skin segmentation but also holds promise for adaptation and expansion to other medical imaging and segmentation tasks.


\bibliographystyle{IEEEtran}
\bibliography{References}

\end{document}